\documentclass[natbib]{svjour3}                     
\smartqed  
\usepackage{graphicx}
 \usepackage{aps-bibstyle}  
\newcommand{\MeV}{\rm{\, MeV }}

\newcommand{\erg}{\rm{\, erg }}

\newcommand{\beq}{\begin{equation}}
\newcommand{\eeq}{\end{equation}}
\newcommand{\ba}{\begin{array}}
\newcommand{\ea}{\end{array}}

\newcommand{\D}{{\mathcal {D}}}

\newcommand{\prd}   {{\it Physical Review D}}

\newcommand{\aj}{\mbox{\it Astronomical J.}}
\newcommand{\apj}{\mbox{\it Astrophys. J.}}
\newcommand{\apjl}{\mbox{\it Astrophys. J.}}
\newcommand{\apjs}{\mbox{\it Astrophys. J.}}
\newcommand{\aap}{\mbox{\it Astron. Astrophys.}}
\newcommand{\araa}{\mbox{\it Annu. Rev. Astron. Astrophys.}}

\newcommand{\jcap}{\mbox{\it J. Cosmol. Astropart. Phys.}}
\newcommand{\mnras}{\mbox{\it Mon. Not. R. Astron. Soc.}}
\newcommand{\nar}{\mbox{\it New Astronomy Rev.}}
\newcommand{\nat}{\mbox{\it Nature}}
\newcommand{\na}{\mbox{\it New A.}}

\newcommand{\physrep}{\mbox{\it Phys. Rep.}}
\newcommand{\pasp}{\mbox{\it Publ. Astr. Soc. Pacific}}
\newcommand{\planss}{\mbox{\it Planetary and Space Science}}
\newcommand{\ssr}{\mbox{\it Space Science Rev.}}
\def\lsim{\raisebox{-0.3ex}{\mbox{$\stackrel{<}{_\sim} \,$}}}
 \def\gsim{\raisebox{-0.3ex}{\mbox{$\stackrel{>}{_\sim} \,$}}}

\begin{document}

\title{Energetic and Broad Band Spectral Distribution of Emission from Astronomical Jets
}


\author{Asaf Pe'er 
}


\institute{A. Pe'er \at
              Physics Department, University College Cork, Cork, Ireland \\
              Tel.: +353-21-4902594\\
              \email{a.peer@ucc.ie}          
}

\date{Received: date / Accepted: date}

\maketitle

\begin{abstract}
Emission from astronomical jets extend over the entire spectral band:
from radio to the TeV $\gamma$-rays. This implies that various
radiative processes are taking place in different regions along
jets. Understanding the origin of the emission is crucial in
understanding the physical conditions inside jets, as well as basic
physical questions such as jet launching mechanism, particle
acceleration and jet composition.  In this chapter I discuss
various radiative mechanisms, focusing on jets in active galactic
nuclei (AGN) and X-ray binaries (XRB) environment.  I discuss various
models in use in interpreting the data, and the insights they provide.
\end{abstract}

\section{Introduction}
\label{intro}

Jets and outflows are very ubiquitous in astrophysics. They are observed in both
galactic objects such as X-ray emitting binaries (XRBs) [For reviews,
  see, e.g., \cite{Fender06, Fender10, Gallo10, Markoff10,
    Maccarone12}], as well as extra-galactic sources, such as active
galactic nuclei (AGNs) [\cite{BBR84, UP95, HK06, Marscher09, Ghis12}],
and on a much smaller scale, gamma-ray bursts (GRBs) [e.g.,
  \cite{LE93, Piran04, Meszaros06}].  Recently, the existence of jet was
inferred in a tidal disruption event (TDE) of a stray star passing
near a massive black hole [\cite{Burrows+11, Levan+11}]. Emission from
jets in both galactic and extra-galactic sources is observed over the
entire spectral range: from radio to the highest $\gamma$-rays, at TeV
energies. In addition to the spectral information, in galactic sources
(X-ray binaries) as well as in jets from GRBs and TDEs, a wealth of
temporal information exists.  Similarly, in the high energy (GeV -
TeV) emission from AGNs, flaring activities on time scales shorter
than $\sim 1$~h has been observed [\cite{Kniffen+93, Buckley+96,
    aharonian+07, albert+07, Aleksic+11}]. Moreover, in nearby jets
from AGNs, such as Cen A, spatial information about emission from
different region along the jet exist [\cite{Hardcastle+09} and
  references therein].

While a wealth of data has existed for several decades now, a detailed
theoretical understanding of emission from jets is still lacking. One
potential explanation is that most work to date has focused on
emission from the accreting (inflow) material, and only in the past
decade or so have more advanced models of emission from the outflow
(jets) emerged.  One notable exception is the emission from GRB jets:
as our understanding of these objects relies nearly entirely on
studying emission from their jets, the theory of emission from GRB
jets is likely the most advanced one to date.  A second reason is the
enormous complexity of these systems. As will be discussed here,
although the nature of the radiative processes is well understood, as
the physical conditions inside and in the vicinity of the jets are
poorly constrained, the data can be interpreted in more than one
way. As a result, a plethora of models exist, and a conclusive picture
is still lacking.

Although different objects share the common property of having jets,
there is a huge difference in scales of the observed objects. While
XRBs and GRBs are stellar-size objects, with a typical mass of the
central BH of $\sim$~few~$M_\odot$, the black holes in the center of
AGNs have masses of $10^6 - 10^9 \, M_\odot$. This difference in
scaling results in very different scales of the resulting jets. In
galactic XRBs the inferred size of the observed jets is typically
100's of AU's ($\gsim 10^{14} - 10^{15}$~cm) [\cite{MJ+12}] while
radio 'blobs' are seen on much larger, sub-pc scales ($\sim
10^{17}$~cm). In GRBs, the jet does not deposit most of its energy in
the environment before reaching $\sim 10^{18}$~cm [\cite{MR97, WRM97}],
although analysis shows that emission exists from the photosphere at
$\sim 10^{12}$~cm [\cite{Axelsson+12}]. Sizes of AGN jets extend to
much larger scales, with giant radio lobes extending to hundreds of
kpc, $\sim 10^{23}$~cm [\cite{Alvarez+00} and references therein].

This difference in scaling implies that the physical conditions, and
hence the leading radiative mechanisms inside the jets, vary with
distance. Nonetheless, the basic emission processes are the same in
all sources.  The leading radiative processes include synchrotron
emission, synchrotron self-Compton and Compton scattering of photons
external to the jet - either photons originating from the accretion
disk, companion star (in XRBs) or from the cosmic microwave (or
infra-red) background (CMB).  If hadrons (mainly protons) are
accelerated to high energies in jets, they can also make a significant
contribution to the emission, particularly at high energies (X and
$\gamma$-ray bands). This contribution is both by direct emission(
e.g., synchrotron), and indirectly, by interacting with photons and
protons to produce secondaries (pions, Kaons and electron-positron
pairs) which contribute to the emitted spectra.  In addition, emission
from the photosphere, defined here as emission that originates from
regions in space in which the optical depth of photons to reach the
observer is larger than unity, may play an important role.

In addition to the spectral analysis, two very important sources of
information exist. The first is temporal variability which provides
strong constraints on the physical conditions inside the jets, and
hence on the emission processes. This played a crucial role in the
development of the leading theory of emission from GRB jets (the
``fireball'' model). Studying the correlated variability seen in the
emission at different wavelengths (radio /IR/optic and
X/$\gamma$-rays) in XRBs is likely the key to understanding of the
emission from these objects [See \cite{Uttley+11}, and the chapters by
  Gallo and Casella in this book]. The second source of information is
spatial analysis, which is particularly useful when studying the
largest-scale jets in AGNs. The ``hot spots'' frequently seen imply
that the physical conditions and the radiative processes vary along
the jet. Thus, a full physical picture must take into account first
the {\it dynamics} of the outflow and second the {\it
  radiation}. Clearly, both parts are connected, and, in addition,
give information about the jet launching process and the properties of
the inner accretion disk.

Another important factor that needs to be considered in analyzing the
emission is {\it geometry}. There are several aspects to this
issue. First, as astronomical jets are mildly relativistic in XRBs
($\Gamma \gsim few$), often relativistic in AGNs ($\Gamma \gsim 10$)
and highly relativistic in GRBs ($\Gamma \gsim 100 - 1000$),
relativistic Doppler effect is important when analyzing the spectrum,
as the jets will rarely point directly towards us.  Emission from
relativistically expanding blobs can lead to an apparent motion faster
than the speed of light (a phenomenon known as ``superluminal motion''
[\cite{Rees66, MiRo94}]).  Second, jets, by definition, have spatial
structure (often referred to as ``structured jets''): a velocity
profile exists, namely $v = v(r, \theta, \phi)$, where the angles
$\theta, \phi$ are measured relative to the jet axis. Thus, a velocity
gradient in the transverse direction (perpendicular to the jet
propagation direction) exists, with an obvious effect on the
scattering between electrons and photons, and hence on the observed
spectra.

Finally, the velocity structure in the radial direction can lead to
confusing definition of jets.  One possibility is that the outflow is
continuous (generating a smooth velocity gradient in the radial
direction) in which case it will be seen as a continuous
jet. Alternatively, the outflow may be fragmented: in this case, the
outflow will be observed as 'blobs' that propagate outward, while
expanding (possibly, but not necessarily, adiabatically). Of course,
the observed emission from these blobs imply that the conditions
inside the blobs are different than those outside. Thus, when studying
emission from these blobs one needs to consider the conditions both
inside the blobs and in the surrounding material. In this chapter, we
will treat both emission from continuous jets as well as from the
blobs.

Thus, a full description of the emission requires understanding of (1)
the dynamics, (2) the geometry and (3) the various radiative
processes. Clearly, I cannot possibly cover the entire physics of jet
emission in one chapter. I will thus focus on key radiative
processes. I will show how the basic, well-known radiative processes
can lead to the wealth of spectra observed. I will also try to point
to basic, unsolved questions which naturally arise when analyzing the
emission. The discussion will be focused on the jet emission from XRB
and AGN environments, which show several similar key properties,
although having different scales. Clearly, many of the
results are relevant to jets in GRBs and TDEs as well.

\section{Basic radiative processes: synchrotron emission}
\label{sec:rad1}

Variable radio emission in AGNs and XRBs is conventionally interpreted
as synchrotron radiation from a non-thermal distribution of
relativistic electrons. Indeed, synchrotron emission, being perhaps
the most straightforward emission mechanism for explaining non-thermal
radiation has been extensively studied since the 1960's [\cite{GS65, BG70}].
Two basic ingredients are needed: energetic particles and a strong
magnetic field. 

Consider a source at redshift $z$ which is moving
at velocity $\beta \equiv v/c$ (corresponding Lorentz factor
$\Gamma = (1-\beta)^{-1/2}$) at angle $\theta$ with respect to the
observer. The emitted photons are thus seen with a Doppler boost $\D =
[\Gamma(1-\beta \cos\theta)]^{-1}$. Synchrotron emission from electrons
having random Lorentz factor $\gamma_{el}$ in a magnetic field $B$ (all
in the comoving frame) is observed at a typical energy
\beq 
\varepsilon_m^{ob} = {3 \over 2} \hbar {q B \over m_e c}
\gamma_{el}^2 {\D \over (1 +z)} = 1.75 \times 10^{-19} B \gamma_{el}^2
      {\D \over (1+ z)} \erg.
\label{eq:nu_m}
\eeq

Thus, when studying this emission, the basic physical questions are: 
\begin{enumerate}
\item What is the origin of the magnetic field ?
\item What is the mechanism that accelerates particles to high
  energies ?  Does this mechanism accelerate only electrons? Are
  protons being accelerated similarly, thereby contributing
  to the emission ?  What is the resulting energy
  distribution of the energetic particles, $n(E) dE$ ?
\item What is the spatial / temporal evolution of the magnetic field
  and particle distribution in different regions along the jet ?
\end{enumerate}

While significant progress has been made in the last few decades,
proper understanding of any of these issues remains elusive.  These
questions are deeply related to the physics of the jet launch
mechanism, and jet composition. While there is no direct observational
test that addresses any of these phenomena, very considerable
theoretical effort supported by state of the art numerical
simulations, as well as indirect interpretation of existing data, all
suggest that these questions are likely intimately related to each
other. We discuss these questions in what follows.

{\bf Origin of magnetic field.} Although the question of magnetic
field generation is a fundamental one, little is known about the exact
mechanism at work in these objects. Roughly speaking, there are two
main (separated) sources of strong magnetic fields: the first is related to
the accretion flow and the jet launching process. Possibly, even if
the magnetic fields do not carry a large fraction of the kinetic
energy, they may still play a key role in jets collimation.  The
second, independent source is magnetic field generation in shock waves
that exist inside the outflow itself (assuming it is irregular), or
when the outflow interacts with its surroundings - the interstellar
medium (ISM) or intergalactic medium (IGM).  A third possible source is
amplification of ISM or IGM magnetic fields when compressed by the
expanding shock waves, but in this case the amplified fields can at most explain
the observed emission in the interaction of the outflow with its
surroundings - they are much too weak to be consistent with the ones
required to explain the observed properties in the inner jet regions.  For a
recent review on magnetic fields in astrophysical jets, see
\cite{PHG12}.

The two leading mechanisms believed to operate for jet launching are
the \cite{BZ77} and \cite{BP82} mechanisms. In the \cite{BZ77}
mechanism, the source of energy is the rotational energy extracted
from a rotating black hole, embedded in a strong magnetic field. The
field itself must be anchored into the accretion flow [\cite{Livio+99,
    Meier01}].  In the \cite{BP82} mechanism, energy is extracted from
the accretion disk by magnetic field lines that leave the disk surface
and extend to large distances. This is accompanied by
centrifugally-driven outflow of material from the inner parts of the
disk, that is attached to the field lines [for further explanation see
  \cite{Spruit10}, as well as the chapter on jet acceleration in this
  book]. Both mechanisms require a strong magnetic field attached to the
disk. At larger distances along the jet, the magnetic field decays as
Poynting flux is conserved.

These ideas have been recently tested and validated with state of the art
numerical GR-MHD simulations [\cite{MKU01, MG04, Mckinney05,
    McKinney06, TNM10, TNM11}]. These models imply that the
magnetic field originates in the inner parts of the disk. The inner
parts of the jets are strongly magnetized (Poynting-flux dominated),
and the magnetic energy is gradually dissipated along the jet. The
dissipated energy is then used to accelerate the particles along the
jet [see, e.g., \cite{VK04, Komissarov+07}]. Recent observations on
parsec-scale in AGNs indicate magnetic field strengths consistent with
those expected from theoretical models of magnetically powered jets
[\cite{OG09}]. However, the picture is far more complicated, since
modeling the broad-band emission (radio - X-rays) on a $\gsim$ parsec
scale from several AGNs show that the magnetic field must be
sub-dominant, and most of the kinetic energy is carried by protons
(particle-dominated jets) [\cite{CF93, KCA02, KTK02, CG08}]. The
mechanism in which magnetic-dominated outflow at the core becomes
particle dominated at larger distances is far from being clear.

Independent of the question of jet launching, a second source of
strong magnetic fields are shock waves that develop as a result of
instabilities within the outflow. These shock waves can result, e.g.,
from fluctuations in the ejection process itself: if a slower moving
plasma shell (or ``blob'') is followed by a faster moving one, the two
shells will eventually collide, producing a pair of forward and
reverse shock waves propagating into each of these blobs. These shock
waves may generate strong magnetic fields by two-stream instabilities
[\cite{Weibel59, ML99}]. In recent years, advances in particle-in cell
(PIC) simulations enabled to study this process by tracing the
instability growth modes [\cite{Silva+03, Frederiksen+04,
    Nishikawa+05, Spit08a}]. The results of these works have
demonstrated that strong magnetic fields are indeed created in
collisionless shock waves.

The key question though, is the decay length of the magnetic field:
the results discussed above also show that the generated field decays
on a very short length scale, of the order of few hundred skin depths
[\cite{Spit08a}]. As observations imply that the emitting region is
many orders of magnitude larger than this scale, there must be a
mechanism that maintains a strong magnetic field extending to much
larger scales. One suggestion is that the amplification of the
magnetic field may be closely related to the acceleration of particles
to high energies [\cite{Keshet+09}]. Thus, while initially the
magnetic field may occupy only a small region close to the shock
front, over time, as particles are accelerated to increasingly higher
energies, the magnetized region expands. This suggestion is difficult
to directly test, due to the numerical complexity of the
problem. Another suggestion is that, due to the turbulent nature of
the post-shock outflow, the magnetic field is maintained over a long
distance behind the shock front [\cite{ZM12}].  Thus, while it is
clear that magnetic fields can be generated in shock waves, the exact
scaling (strength and decay length) of these fields is still a matter
for debate.

{\bf Particle acceleration}. It should be emphasized that the
existence of cosmic rays, [charged particles that are observed at
energies as high as $\gsim 10^{20}$~eV; for a recent review, see
  \cite{KO11}], is a direct evidence that particle acceleration to
ultra-high energies takes place in astronomical objects. However,
there is no direct information on the exact nature of the cosmic ray
sources, nor on the nature of the acceleration process
itself. Hence the question of lepton (electrons and positrons)
acceleration is inferred indirectly, by fitting the observed spectra
from various objects. It is most likely that acceleration takes place
in several different locations: in the nucleus, in the hot spots and
possibly additional locations along the jet axis.

The most widely discussed mechanism for acceleration of particles is the
{\it Fermi mechanism} [\cite{Fermi49, Fermi54}], which requires the
particles to cross back and forth a shock wave.  A basic explanation of
this mechanism can be found in the textbook by \cite{Longair11}. For
reviews see \cite{Bell78, BO78, BE87, JE91}. In this process, the
accelerated particle crosses the shock multiple times, and in each crossing
its energy increases by a (nearly) constant fraction, $\Delta E / E
\sim 1$. This results in a power law distribution of the accelerated
particles, $N(E) \propto E^{-S}$ with power law index $S \approx 2.0 -
2.4$ [\cite{Kirk+98, Kirk+00, Ellison+90, Achterberg+01, ED04}].
Recent developments in particle-in-cell (PIC) simulations have allowed 
to model this process from first principles, and study
it in more detail [\cite{Silva+03, Nishikawa+03, Spit08, SS09b,
    Haug11}]. However, due to the numerical complexity of the problem,
these simulations can only cover a tiny fraction ($\sim
10^{-8}$) of the actual emitting region in which energetic particles
exist. Thus, these simulations can only serve as guidelines, and
the problem is still far from being fully resolved. Regardless of the
exact details, it is clear that particle acceleration via the Fermi
mechanism requires the existence of shock waves, and is thus directly
related to the internal dynamics of the gas inside the jet, and
possibly to the generation of magnetic fields, as mentioned above.

An alternative model for particle acceleration is magnetic
reconnection. The basic idea is that when magnetic field lines change
their topology and form a reconnection layer, magnetic energy is
released. Part of the generated energy may be used to accelerate
particles to high energies [\cite{RL92, Lyutikov03, LL08, Lazarian+11,
    MU12}]. This idea is very appealing if jets are highly magnetized
(at least close to the core), as is suggested by the leading theories
of jet launching. In fact, it is not clear that the conditions that
enable particle acceleration to high energies in shock waves exist at
all in highly magnetized outflows [\cite{SS09b, SS11}], in which case
the Fermi mechanism may not be at work.  However, theoretical
understanding of this process, and its details (e.g., what fraction of
the reconnected energy is being used in accelerating particles, or the
energy distribution of the accelerated particles) is still very
limited.

Although the power law distribution of particles resulting from
Fermi-type, or perhaps magnetic-reconnection acceleration is the most
widely discussed, we point out that alternative models exist. One such
model involves particle acceleration by a strong electromagnetic
potential, which can exceed $10^{20}$~eV close to the jet core
[\cite{Lovelace76, Blandford76, Neronov+09}]. The accelerated
particles may produce a high energy cascade of electron-positron
pairs. Additional model involves stochastic acceleration of particles
due to resonant interactions with plasma waves in the black hole
magnetosphere [\cite{Dermer+96}].

Several authors have also considered the possibility that particles in
fact have a relativistic quasi-Maxwellian distribution [\cite{JH79,
    CJ80, WZ00, PC09}]. Such a distribution, with the required
temperature ($\sim 10^{11} - 10^{12}$~K) may be generated if
particles are roughly thermalized behind a relativistic strong shock
wave [e.g., \cite{BM77}]. Interestingly, this model is consistent with
several key observations, as will be discussed below.

{\bf Spatial and temporal distributions.}  The uncertainty that exists
in both the origin of the magnetic field as well as the nature of the
particle acceleration process is directly translated to an uncertainty
in the spatial and temporal distributions of these two quantities, and
hence on the emission pattern.  If the magnetic field originates in
the disk, then as the jet expands the magnetic field must decay. For
example, if the cross sectional radius of the jet is $r=r(z)$ where
$z$ is the direction along the jet axis, then Poynting flux
conservation implies $B \propto r^{-1}$. If, on the other hand, the
dominant process for magnetic field generation is two stream
instability in shock waves, the magnetic field then traces the shock
wave location. Thus, strong magnetic fields are expected only above a
certain radius, where plasma shells collide. This is very likely the
case in the spatially resolved ``radio blobs'' seen in XRBs, as well
as in ``knots'' observed along AGN jets.

The magnetic field strength may also be different in the two
possibilities discussed. Lacking a complete theory, it is commonly
assumed that the generated magnetic field carries a constant fraction,
$\epsilon_B$ of the kinetic energy dissipated by the shock wave,
$B^2/8\pi = \epsilon_B U$. Here, $U$ is the (post-shock) energy
density of the plasma. Estimated values for $\epsilon_B$ based on
fitting the data vary from equipartition ($\epsilon_B = 1/3$; [\cite{Miller-Jones+05, Cerruti+13}]) to
$\epsilon_B \sim 10^{-2}$, possibly even lower [\cite{CG08, SBK13}].

The spatial and temporal distribution of the energetic particles is
determined by several factors. Once accelerated to high energies, the
radiating particles lose their energy both adiabatically as the jet
expands, and radiatively, as they radiate their energy. Thus, in order
to understand their spatial distribution, one needs to know (1) the
initial distribution of the energetic particles accelerated by the
acceleration process (determined by the unknown nature of this
process). (2) The dynamics of the plasma; and (3) the physical
conditions inside the plasma, that govern the energy loss rate.

\subsection{Spectral shape: basic considerations and Maxwellian distribution of electrons}

The discussion above points towards high uncertainty in our knowledge of the initial
energy distribution of particles produced by the
acceleration process. It is commonly believed that the acceleration
process produces a power law distribution $n(E) dE \propto E^{-S}$,
with $S \approx 2.0 - 2.4$. This is based on (1) theoretical
expectations from Fermi acceleration, and (2) interpretation of broad
band synchrotron emission. However, a few words of caution are necessary
here. First, as discussed above, it is not clear that the Fermi process is
necessarily the acceleration mechanism at work in these
objects. Second, as was recently shown [\cite{PC09}] and will be
discussed below, the observed data can be interpreted in a way that
does not require a power law distribution of the radiating
particles. Thus, evidence for the existence of a power law distribution is inconclusive.

Even if the acceleration process is indeed Fermi-type in shock waves,
then the resulting power law distribution is expected to be limited to a certain
region in energy space. As particles cross the shock front, they
thermalize. Strong shock waves propagating at Lorentz factor $\Gamma$
into a cold material of density $n$ compress the material so that its
density in the downstream region is $4 \Gamma n$. The material is
being heated: the energy density in the downstream region is $4
\Gamma^2 n m_p c^2$. Thus, the average energy per particle in the
downstream region is $\approx \Gamma m_p c^2$ [\cite{BM76}]. If a
fraction $\epsilon_e \leq 1$ of this energy is carried by the
energetic electrons, then (neglecting a possible contribution from
pairs) the expected Lorentz factor of the electrons is $\gamma_{el}
\approx \Gamma \epsilon_e m_p/m_e$. Note that this is the Lorentz
factor associated with the random motion of the electrons as they
cross the shock front, and should not be confused with the Lorentz
factor associated with the bulk motion of the flow, which is of the
order of $\Gamma$. Thus, even in mildly-relativistic outflows ($\Gamma
\gsim 1$), the electrons in the downstream region may still have
(random) Lorentz factor of few hundreds, provided that $\epsilon_e$ is
close to equipartition.

One can thus conclude that regardless of the question of whether
electrons are accelerated to a power law distribution, they are still
expected to be heated to high energies (high Lorentz factors) when
shock waves exist. Hence, if no further acceleration process are present, the
electrons will have a Maxwellian distribution with typical Lorentz
factor $\gamma_{el} \lsim 10^3$ (assuming $\epsilon_e$ close to
equipartition). In the vicinity of a magnetic field $B$, which could
naturally be generated by the same shock wave, electrons at the peak
of the Maxwellian distribution will emit at a characteristic energy given by
Equation \ref{eq:nu_m}. For typical value $\gamma_{el} \sim 10^3$ and
$B \sim 1$~G, Equation \ref{eq:nu_m} implies a characteristic observed
frequency in the optical band.

This result implies that in order to explain the observed flat radio
spectra seen in many objects, there is no need to invoke a power law
distribution of the accelerated particles. It is enough to consider a
power law decay of the magnetic field along the jet, $B(r) \propto
r^{-\alpha}$ to obtain a power law decay of the peak synchrotron
frequency below the optical band, in accordance to Equation
\ref{eq:nu_m}. A power law spectrum would be observed if the emission
is not spatially resolved, but is integrated over some distance along
the jet along which the magnetic field decays. This is a typical
scenario for the inner parts of AGN jets, as well as for jets in XRBs.

\subsection{Power law distribution of the accelerated particles}

It is possible to envision a different model, in which the energy
distribution of the accelerated electrons is a power law. In fact,
historically this model was the first to be suggested in explaining
the observed spectra [\cite{VdL66, BK79}], and is still the most
widely-discussed one.

An uncertainty lies in the fraction of particles that are being
accelerated: as the electrons cross the shock wave, they have a
thermal distribution with typical Lorentz factor $\gamma_{el}$. As
some fraction continues to cross the shock front multiple times, this
fraction obtain a power law distribution.  What fraction of particles
are accelerated to a power law distribution above the Maxwellian is
unclear. Recent PIC simulations suggest that only a small fraction of
the population, $\epsilon_{pl} \approx 1\% - 10\%$ form a power law
tail at higher energies [\cite{Spit08}]. However, as discussed above,
these conclusions are far from being certain, and it is possible that
the fraction is much higher, perhaps even closer to 100\%.

A theoretical limit on the maximum energy is obtained by the requirement
that the acceleration time must be shorter than the minimum
energy loss time (e.g., due to synchrotron radiation or Compton
scattering) and the time in which the accelerated particle is confined
to the accelerated region. In a plasma which moves relativistically
with Lorentz factor $\Gamma$, the acceleration time in Fermi-type
acceleration is [e.g., \cite{Norman+95}]
\beq
t_{acc} = {\eta E^{\rm
  ob} \over \Gamma Z q B c}. 
\label{eq:t_acc}
\eeq
Here, $Z q$ is the charge of the particle (the same equation holds for
electrons, protons as well as heavy nuclei with $Z$ protons), and
$E^{\rm ob}$ is the energy of the energetic particle in the observer's
frame. The exact value of the dimensionless
factor $\eta \geq 1$ depends on the uncertain details of the
acceleration process: for example, in non-relativistic diffusive shock
acceleration, this factor corresponds to $\eta = (20/3) \beta^{-2}$ in
the Bohm limit for parallel shocks [e.g., \cite{BE87}].

The second requirement constraints the size of the acceleration
region. For typical values of parameters that govern the emission in
XRBs and AGNs, it is not very restrictive. On the other hand, the
requirement that the acceleration time is shorter than the radiative
cooling time puts a stronger constraint on the maximum energy of the
accelerated particles.  The radiative cooling time of energetic
electrons due to synchrotron emission and Compton scattering is
\beq 
t_{\rm cool} = {E \over P} = {\gamma_{el} m_e c^2 \over (4/3) c
  \sigma_T \gamma_{el}^2 u_B (1 + Y)} = {6 \pi m_e c \over \sigma_T
  B^2 \gamma_{el} (1 +Y)},
\label{eq:t_cool}
\eeq
where $u_B \equiv B^2/8\pi$ is the energy density in the magnetic
field, $\sigma_T$ is Thomson's cross section and $Y$ is Compton
parameter. Comparing the radiative cooling time in Equation
\ref{eq:t_cool} to the acceleration time in Equation \ref{eq:t_acc}
gives a theoretical upper limit on the energy of the accelerated electrons,
\beq
\gamma_{\max} = \left( {6\pi q \over \eta B \sigma_T (1+Y)} \right).
\label{eq:gamma_max}
\eeq
Using the derived value of $\gamma_{\max}$ from Equation
\ref{eq:gamma_max} in Equation \ref{eq:nu_m} gives a very interesting
result: the characteristic energy of photons emitted by these
electrons,
\beq
\varepsilon_{\max}^{\rm ob} = 240 {\mathcal{D} \over \eta (1+Y)
  (1+z)}~\MeV
\label{eq:nu_max}
\eeq
is {\it independent on the strength of the magnetic field}.  This
result implies that regardless of the value of the magnetic field, if
indeed particles are accelerated by a Fermi-type mechanism in shock
waves, synchrotron emission is expected to be observed at all
energies up to the $\gamma$-ray band.  Thus, it is possible, at least
from a theoretical perspective, that synchrotron photons have a
significant contribution to the emission at X and $\gamma$-ray energies.

\subsection{Broad band, spatially resolved synchrotron spectrum}

The discussion above implies that even if synchrotron emission is the
only source of radiation, and even if the magnetic field is constant,
the complex distribution of the energetic particles leads to a complex
observed spectrum. In addition to the two frequencies discussed,
$\nu_m = \varepsilon_m/h$ (see Equation \ref{eq:nu_m}) and $\nu_{\max}
= \varepsilon_{\max}/h$ (the later exists only if the acceleration
process produces a power law), there are additional two inherent
characteristic frequencies. The first is the synchrotron self
absorption frequency, $\nu_{SSA}$, below which synchrotron photons are
absorbed.  This frequency can be either above or below $\nu_m$. The
exact value of $\nu_{SSA}$ depends on the magnetic field strength and
the distribution of the radiating particles [for discussion see
  \cite{RL79}].  For typical parameters, $\nu_{SSA} < \nu_m$. However,
given the uncertainty that exists in the acceleration process and the
strength of the magnetic fields, it is possible to envision scenarios
in which $\nu_m < \nu_{SSA}$ .

The fourth frequency is the cooling frequency, $\nu_c$. This is the
characteristic emission frequency from particles whose radiative
cooling time (given by Equation \ref{eq:t_cool}) is equal to the
characteristic plasma expansion time, $t_{dyn} \simeq r/\Gamma c$,
where $r$ is the radius of the expanding plasma. Since the cooling
time is inversely proportional to the particle's Lorentz factor,
$t_{cool} \propto \gamma^{-1}$ (see Equation \ref{eq:t_cool}),
energetic particles cool faster than low energy ones. Above a certain
Lorentz factor, denoted by $\gamma_c$, particles cool faster than the
dynamical time. Thus, if particles are accelerated only once,
$\gamma_c = \gamma_{\max}$. However, if the acceleration continuously
produces a power law distribution of energetic particles, $n(\gamma)
d\gamma \propto \gamma^{-S}$, then $\gamma_c$ marks a transition in
the steady state distribution. By solving the continuity equation, it
is easy to show [see, e.g., \cite{Longair11}] that for $\gamma \gg
\max( \gamma_c , \gamma_m)$, the steady particle distribution is $n(\gamma)
d\gamma \propto \gamma^{-(S+1)}$. If $\gamma_c < \gamma_{el}$, then in
the region $\gamma_c \ll \gamma \ll \gamma_{el}$ the steady
distribution is $n(\gamma) d\gamma \propto \gamma^{-2}$.  This break
in the particle distribution is directly translated to a break in the
emitted spectrum. As the synchrotron spectrum from a power law
distribution of particles with power law index $S$ is $F_\nu \propto
\nu^{-(S-1)/2}$, at frequencies above $\nu_c$, this power law changes
to $F_\nu \propto \nu^{-S/2}$.

Thus, in Fermi-type acceleration, four breaks in the spectrum are
expected (see Table \ref{tab01}). Even if the acceleration mechanism produces only a
Maxwellian distribution of hot particles, at least two break
frequencies ($\nu_{SSA}$ and $\nu_m$) are unavoidable.
\begin{table}[htbp]
\caption{Key break frequencies}
\begin{tabular} {ll} \hline \\
 $\nu_m$ & Synchrotron emission frequency from electrons at typical Lorentz factor $\gamma_{el}$ \\
& (given in Equation \ref{eq:nu_m}).  \\
 $\nu_{ssa}$ & Self absorption frequency; Photons at $\nu< \nu_{ssa}$ are absorbed. \\
 $\nu_c$ & Cooling frequency; \\
&  A break in the spectrum caused by rapid cooling of electrons at high energies. \\
 $\nu_{\max}$ & Maximum emission frequency from Fermi- accelerated electrons; \\
 &  no synchrotron emission is expected at higher frequencies. \\
\hline
\label{tab01}
\end{tabular}
\end{table}

\begin{figure*}
 \includegraphics[width=0.5\textwidth]{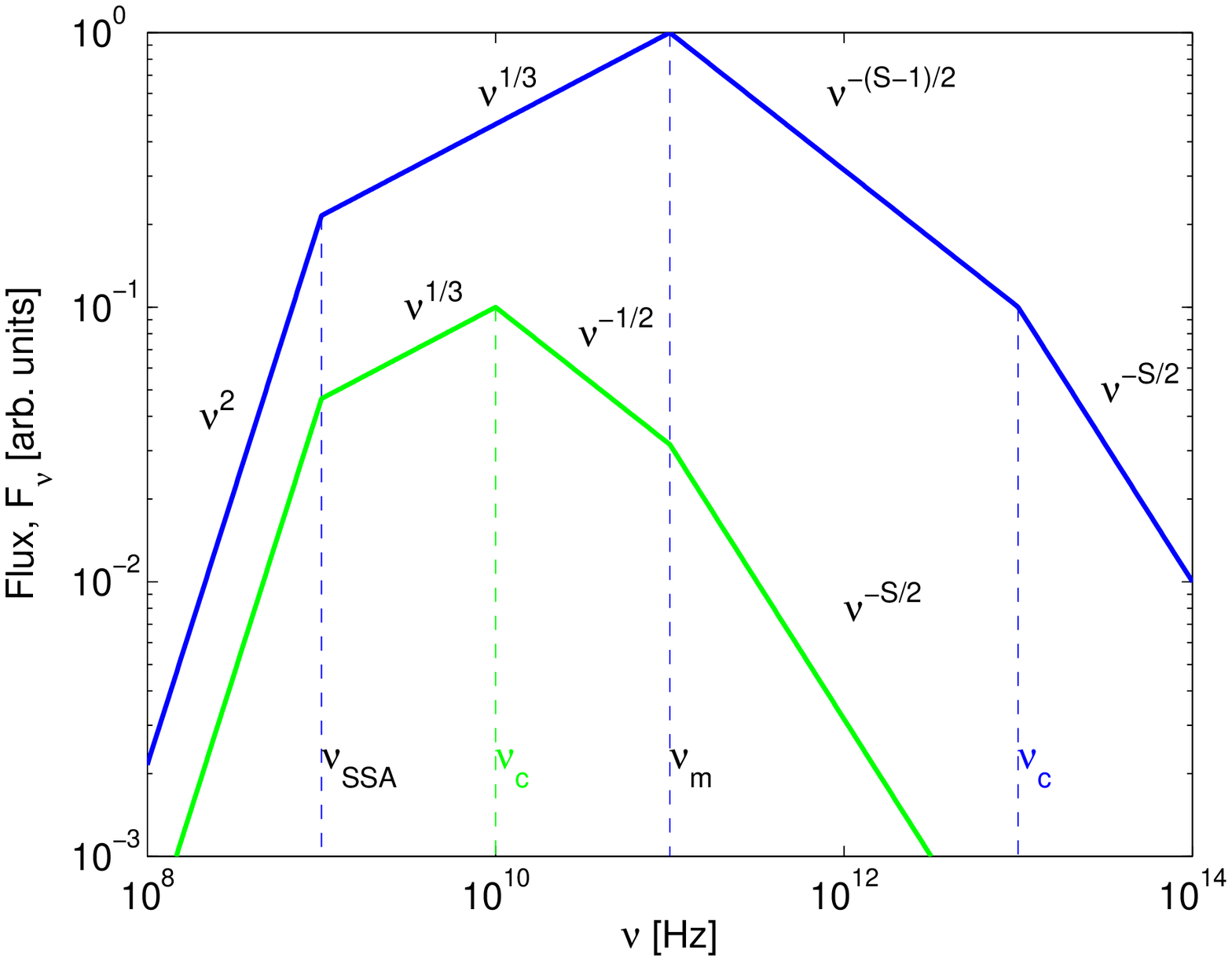}
 \includegraphics[width=0.5\textwidth]{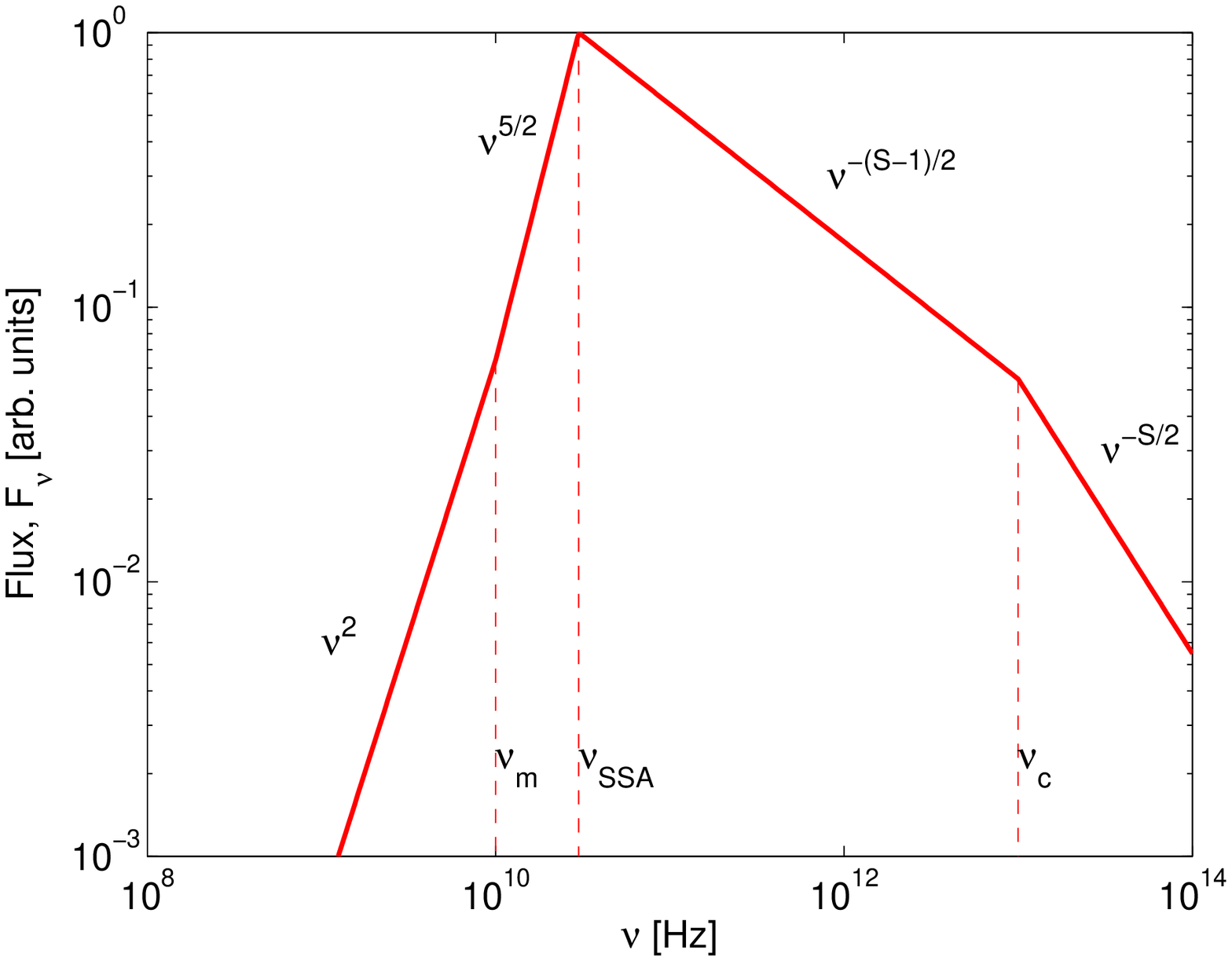}
\caption{Broad band synchrotron spectra from a power law distribution
  of energetic particles in a steady magnetic field, as expected from
  a small jet slab. The flux and the chosen values of the break frequencies
  are arbitrary, and depend on the exact values of the magnetic field,
  particles energies and number of radiating particles.  {\it Left:}
  the spectra expected when $\nu_{SSA} < \nu_m < \nu_c$ (blue) peaks
  at $\nu_m$. In the case $\nu_{SSA} < \nu_c < \nu_m$ (green), the
  spectrum peaks at $\nu_c$. These spectra were considered in the
  model of \cite{PC09}.  {\it Right:} the broad band spectra in the
  scenario $\nu_m < \nu_{SSA} < \nu_c$ peaks at $\nu_{SSA}$. This is
  the scenario considered in the model of \cite{BK79}, and is commonly
  considered in the literature thereafter.  }
\label{fig:spectra}      
\end{figure*}

In the scenario where $\nu_{SSA} < \nu_m < \nu_c$ the peak of the
spectrum occurs at $\nu=\nu_m$. Denoting by $F_{\nu,\max}$ the
observed peak flux, the broad band synchrotron spectrum is [e.g.,
  \cite{MR97, SPN98}]
\beq 
F_\nu = F_{\nu,\max} \times \left\{ 
\ba{lc} 
(\nu/\nu_{SSA})^2 (\nu_{SSA}/\nu_m)^{1/3} & \nu < \nu_{SSA} \\ 
(\nu/\nu_m)^{1/3} & \nu_{SSA} < \nu < \nu_m \\ 
(\nu/\nu_m)^{-(S-1)/2} & \nu_m < \nu < \nu_c \\ 
(\nu_c/\nu_m)^{-(S-1)/2} (\nu/\nu_c)^{-S/2} & \nu_c < \nu < \nu_{\max}
\ea \right.  
\label{eq:spectrum1}.
\eeq 
If, on the other hand $\nu_{SSA}< \nu_c < \nu_m$, the peak of the emission is at
$\nu_c$, and the spectral shape is
\beq 
F_\nu = F_{\nu,\max} \times \left\{ 
\ba{lc} 
(\nu/\nu_{SSA})^2 (\nu_{SSA}/\nu_c)^{1/3} & \nu < \nu_{SSA} \\ 
(\nu/\nu_c)^{1/3} & \nu_{SSA} < \nu < \nu_c \\ 
(\nu/\nu_c)^{-1/2} & \nu_c < \nu < \nu_m \\ 
(\nu_m/\nu_c)^{-1/2} (\nu/\nu_m)^{-S/2}  & \nu_m < \nu < \nu_{\max}
\ea \right.  
\label{eq:spectrum2}.
\eeq 
Finally, the model of \cite{BK79} can be viewed as a model in which
$\nu_m < \nu_{ssa} \ll \nu_c$. In this case, the peak of the emission
is at $\nu_{SSA}$, and the broad band spectrum is
\beq 
F_\nu = F_{\nu,\max} \times \left\{ 
\ba{lc} 
(\nu/\nu_{m})^2 (\nu_{m}/\nu_{SSA})^{5/2} & \nu < \nu_{m} \\ 
(\nu/\nu_{SSA})^{5/2} & \nu_{m} < \nu < \nu_{SSA} \\ 
(\nu/\nu_{SSA})^{-(S-1)/2} & \nu_{SSA} < \nu < \nu_c \\ 
(\nu/\nu_c)^{-S/2} (\nu_c/\nu_{SSA})^{-(S-1)/2}  & \nu_c < \nu < \nu_{\max}
\ea \right.  
\label{eq:spectrum3}.
\eeq 
These spectra are shown in Figure \ref{fig:spectra}. 

\subsection{Integrated spectrum: flat radio emission}
 
The broad band spectrum considered above is developed under the
assumptions that the radiating particles have a power law distribution
and that the magnetic field is steady.  In reality, once accelerated,
or even during the acceleration the energetic particles propagate
along the jet. As the magnetic field strength varies along the jet,
the break frequencies - $\nu_{SSA},~ \nu_m$ and $\nu_c$ (but not
$\nu_{\max}$!) are different in different regions along the jet.  If
the jet is spatially resolved, this implies that different regions
along the jet are characterized by different spectra.  If the jet is
spatially unresolved, as is the case in XRBs and the inner parts of
jets in AGNs, then the observed spectrum is obtained by integrating
over different emission regions, each characterized by different break
frequencies. This integration naturally leads to the observed power law
spectra, such as the flat spectra frequently observed at radio
frequencies, in both AGNs (the so called ``flat spectrum radio
quasars'', or FSRQ), and more recently in XRBs [\cite{Hynes+00,
    Fender01, Fender06}, and chapters in this book by Fender, Gallo and Casella].

\begin{figure*}
 \includegraphics[width=0.9\textwidth]{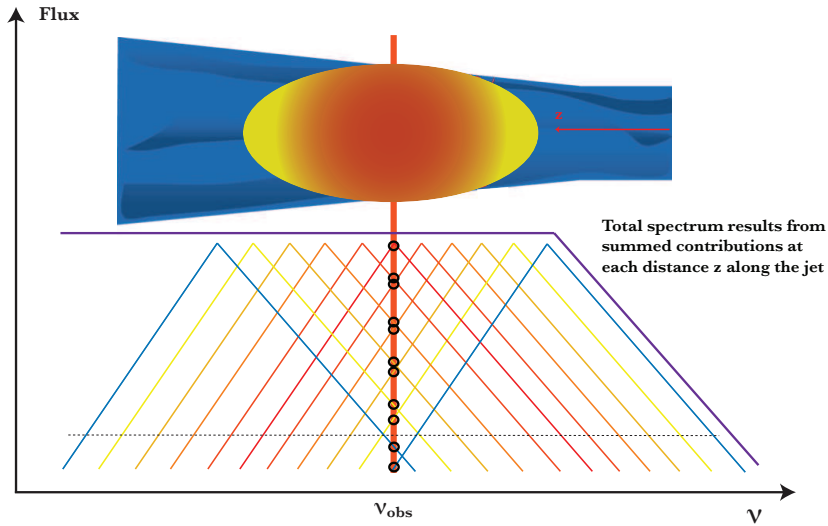}
\caption{Along the jet, the magnetic field decays. Thus, as the
  radiating particles propagate along the jet, the break frequencies
  presented in Figure \ref{fig:spectra} decay. While the emission from
  each slab has the same spectrum as presented in Figure
  \ref{fig:spectra} (right) peaking at $\nu_{SSA}$ (thin lines), when
  integrated over a spatially unresolved region along the jet, the observed
  radio spectrum is flat. The cartoon here is taken from
  \cite{Markoff10}.}
\label{fig:flat1}      
\end{figure*}

If the origin of the magnetic field is in the disk, it is expected to
decay along the jet as a power law with distance. This decay, which
results in a corresponding decrease of the break frequencies, is all that
is needed to produce a power law spectrum in the radio band, in
particular the flat spectrum that is typically observed. This was first noted by
\cite{BK79}. In this model, a conical jet with $B(r) \propto r^{-1}$
and steady outflow velocity resulting in a number density variation along
the jet $n(r) \propto r^{-2}$ was analyzed. Only the evolution of
the self absorption frequency, $\nu_{SSA}$ along the jet was
considered.  In the more general framework considered here, this is
equivalent to a model in which particles are accelerated to a power
law energy distribution with $\nu_m < \nu_{SSA} \ll \nu_c$ as is
presented in Figure \ref{fig:spectra} (right).  In this scenario, the
emission from a jet slab (in which the magnetic field is constant),
peaks at $\nu_{SSA}$. It is straightforward to show [see, e.g.,
  \cite{RL79, BK79}] that these conditions lead to a decay of the self
absorption frequency along the jet, $\nu_{SSA} \propto r^{-1}$, while
the flux from a slab along the jet axis at the self absorption frequency is constant,
$dF_\nu|_{\nu_{SSA}} \propto \nu^0$. Thus, when integrated over an
unresolved distance along the jet a flat radio spectrum is
obtained. This is demonstrated in Figure \ref{fig:flat1}, taken from
\cite{Markoff10}.

This basic idea was extended by several authors in various
aspects. For example, \cite{Marscher80} considered different viewing
angles, while \cite{Reynolds82} considered different dynamical models
for the outflow. In another work, \cite{HJ88} considered a more
refined jet geometry, as well as adiabatic (though, not radiative)
energy losses.  Several works (\cite{FB95, LB96, FB99, HS03, BRP06}) have connected the jet
properties to the disk properties, and refined the inner jet
dynamics. This dynamics was used by several authors [\cite{Markoff+01,
    Markoff+03, Markoff+05, YCN05, YC05, Maitra+09}] to model the
broad band spectra of XTE J1118+480 and GX 339-4. These works included a
self consistent modeling of the emission from the radio all the way to
the X-ray band. In modeling the X-ray emission, the power law
assumption was used to fit both the synchrotron emission and the
synchrotron self Compton (SSC) emission [\cite{Markoff+01, Gallo+07,
    Migliari+07, Maitra+09}; see further discussion below].

While these models show significant improvement in treating the
dynamical properties of jets, the basic radiative mechanism discussed
by \cite{BK79} remains key to all of them.  The radiative particles
are assumed to have a power law distribution, and the peak of the
emission is at $\nu_{SSA}$. The decrease of $\nu_{SSA}$ along the jet
due to the decay of the magnetic field is the origin of the flat radio
spectra.

\begin{figure*}
 \includegraphics[width=0.9\textwidth]{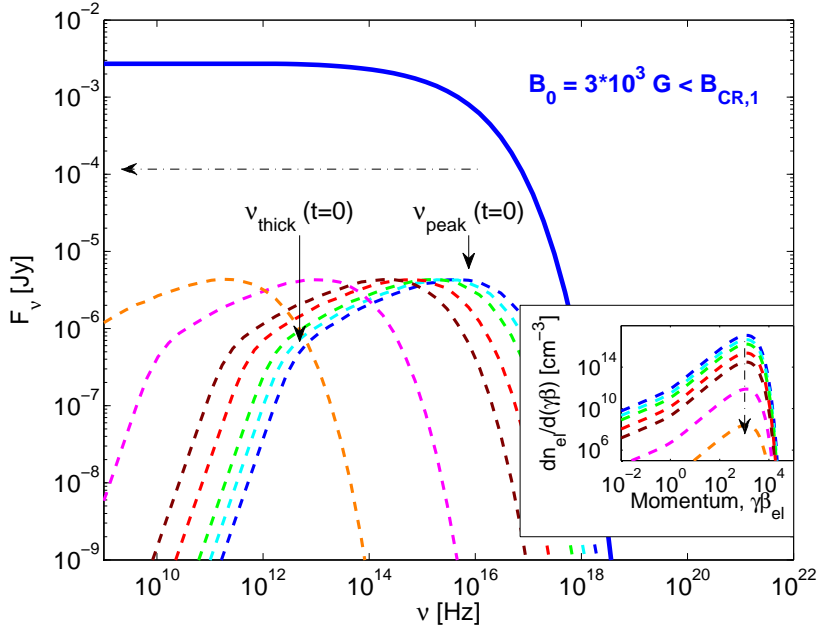}
\caption{The model of \cite{PC09} provides an alternative way to
  explain flat radio spectra without the need for a power law
  distribution of the energetic particles. Emission from a Maxwellian
  distribution (dashed lines) is characterized by two break
  frequencies, $\nu_m$ (described in the figure as $\nu_{peak}$) and
  $\nu_{SSA}$ (or $\nu_{thick}$). In the model considered, $\nu_m >
    \nu_{SSA}$, and thus the emission from a single slab peaks at
    $\nu_m$. As the particles propagate along the jet $\nu_m$ decays,
    and thus the integrated spectra resulting from a Maxwellian distribution of
    particles (inner set) is flat (thick blue line). The value of
    the magnetic field, $B = 10^{3.5}$~G is arbitrarily chosen for
      demonstration purposes only.}
\label{fig:flat2}      
\end{figure*}

An alternative approach was suggested by \cite{PC09}. Based on the
idea of a single acceleration episode and the inclusion of particle
cooling first proposed by \cite{Kaiser06}, this model considers a
scenario in which $\nu_{SSA} < \nu_m$. Thus, the peak of the emission
from a given 'slab' is at $\nu_m$ (or $\nu_c$) rather than at
$\nu_{SSA}$. As was shown in this work, as a result of the decaying
magnetic field along the jet, the decay law of $\nu_m$ is identical to
the decay law of $\nu_{SSA}$. Thus, a flat radio spectrum is naturally
obtained in an analogous way to the \cite{BK79} model, by integrating
over emitting regions inside the jet (see Figure \ref{fig:flat2}).

A conceptual difference between this model and the \cite{BK79} model
is that the former does {\it not} require a power law distribution of
the accelerating electrons. As $\nu_m$ is one of the natural
frequencies obtained from a Maxwellian distribution of radiating
particles, a flat radio spectrum is obtained even if the acceleration
process does not accelerate particles to a power law distribution. A
second difference is that the change in particle distribution due to
cooling is inherently considered. Thus, in a region of strong magnetic
field the radiating particles rapidly cool, $\nu_c < \nu_m$, and the
flux $F_\nu \propto \nu^{-1/2}$ (see Equation
\ref{eq:spectrum2}). This result is independent of the details of the
acceleration process and the magnetic field structure.  Finally, if
the magnetic field is very strong, rapid cooling of the particles as
they propagate along the jet leads to absorption of the emission peak,
$\nu_m < \nu_{SSA}$ for emission that occurs above a certain
radius. This, in turn, results in a suppression of the radio emission,
as seen in several objects [\cite{CP09}].

One assumption common to all jet models is that the decay of the
magnetic field leads to a decay of the characteristic frequencies
along the jet. Thus, although unresolved, emission at low radio
frequencies is expected from distant regions along the jet. Conversely,
emission at higher energies -  microwave, optical X- and $\gamma$-rays - 
must originate from the inner parts of the jet, where the
discrimination between the outflow (jet) and inflow (accretion) may be
very difficult. As shown in Equation \ref{eq:nu_max}, at least from a
theoretical perspective synchrotron photons can be observed
nearly to the GeV-range, irrespective of the strength of the magnetic
field.

\subsection{Fragmented outflow: emission from radio blobs}

The flat radio emission is a natural outcome of emission in a power
law decay of the magnetic field along the jet. Thus, the above
discussed models are relevant when the outflow is continuous. However,
fragmented ejection of material, in the form of radio ``blobs'' are
frequently observed in both XRBs, such as GRS1915+15
[\cite{MiRo94, RM99, Fender+99, Miller-Jones+05}] as well as AGNs.

The different emission observed from these blobs with respect to their
environment indicates different physical conditions inside the
blobs. Particularly, it implies that the magnetic field and/or the
distribution of radiating particles inside the blobs is different than
outside of them. This is a natural consequence if the magnetic field and
particle acceleration originate in shock waves, as discussed above. As
the blobs propagate outward they expand. The expansion can be
adiabatic, but not necessarily (it could be confined by, say an
external magnetic field). Thus, one can deduce scaling laws for the
evolution of the magnetic field and the particle distribution inside
the expanding blobs. The basic model was suggested by \cite{VdL66},
and extended by \cite{HJ88, AA99}.

The key radiative model is similar to the \cite{BK79} model, namely
that particles are accelerated to a power law distribution, and the
emission peaks at $\nu_{SSA}$. However, the scaling laws are
different. The basic assumption is that particles do not enter or
leave a blob, which is adiabatically expanding. Conservation of
magnetic flux implies a decay of the magnetic field $B \propto
L^{-2}$, and adiabatic cooling implies a decline in the particles'
energy, $\gamma \propto L^{-1}$, where $L$ is the comoving size of the
expanding blob. Since the emitted frequency $\nu \propto B
\gamma_{el}^2$ (see Equation \ref{eq:nu_m}), one can derive the
scaling law of the Lorentz factor of the radiating electrons at
observed frequency $\nu$ to be $\gamma_{el} \propto \nu^{1/2} B^{-1/2}
\propto L$. For power law distribution $N(\gamma) d \gamma = k
\gamma^{-S}$, the synchrotron flux then scales as $F_\nu \propto k
B^{(S+1)/2} L^3$, where $k \propto L^{(-S+2)}$ [see \cite{VdL66,
    RL79}].

These scaling laws thus give a testable prediction, $F_\nu \propto
L^{-2S} \sim t^{-2S}$. When confronted with observations
[\cite{Rod+95}], the observed decline is not as steep as the
theoretical prediction. Thus, the simplified version of the theory
needs to be adjusted. One natural possibility is that the expansion is
not adiabatic. For example, reverse shock may play a significant role
in determining the evolution of these blobs [R. Narayan, private
  communication].

\section{Compton scattering and the origin of the X-ray spectrum}

While there is a consensus that the radio spectrum originates from
synchrotron radiation (although the full details of the process are
uncertain), the origin of the X-ray and $\gamma$-ray emission is far
more debatable. As shown above, synchrotron emission can extend up to
hundreds of MeV.  However, at these energies, there are alternative
sources of emission.  In particular, Compton scattering of low energy
photons by energetic electrons is a natural, alternative way to
produce emission at these bands. Due to the larger cross section, even
if hadrons (protons) contribute to the emission, their contribution to
IC scattering is expected to be negligible compared to the electrons
contribution.

Energetic electrons radiate their energy via both synchrotron
radiation and IC scattering.  The total power emitted by IC process
is [\cite{RL79}]
\beq
{P_{IC} \over P_{syn}} = {U_{ph} \over U_{B}}, 
\eeq
where $P_{syn}$ is the synchrotron power, $U_B$ and $U_{ph}$ are the
energy densities in the magnetic and photon fields,
respectively. Thus, if $U_{ph} > U_B$, most of the electrons' energy
is radiated by IC scattering rather than synchrotron. However, even if
$U_B > U_{ph}$, it is still possible that IC scattering is the main source
of emission at a given frequency band.

In understanding Compton scattering, the basic questions are therefore:
\begin{enumerate}
\item What is the origin and spectral distribution of the energetic
  electrons ? Obviously, this is a similar question to the one that
  lies in the heart of understanding synchrotron emission, as the same
  electrons radiate both synchrotron photons and IC photons.
\item What is the origin of the upscattered photon field ? Do these
  photons originate inside the jet (e.g., by synchrotron emission), or
  are they external to the jet (e.g., originating in the accretion flow or CMB)
  ?
\item Since electrons in the inner parts of the accretion flow are hot enough to
  emit in the X-ray band, is there a simple way to discriminate disk
  and jet emission by observing at this band ?
\end{enumerate}

The third question is particularly puzzling, and is the source of an
intense debate.  As discussed above, synchrotron emission from the
inner parts of the jet, where the magnetic field is strongest, are
expected to contribute to the observed flux at the X- and possibly
also $\gamma$-ray frequencies. These regions are close to the inner
parts of the inflow. Thus, discriminating between the inflow and jet as
the sources of X-ray radiation is very challenging.

While IC emission from particles in AGN jets is well established, most
works on X-ray emission in XRBs are focused on IC emission from the
inner parts of the accretion flow. A few notable works are by \cite{ST80, HM93,
  Titarchuk94, MZ95, Esin+97, Poutanen98, CB+06, Yuan+07}, to name
only a handful. As this is the subject of a separate chapter in this book,
we will not discuss it here.  I will point though, that there are
various reasons to consider IC scattering from electrons inside the jets in
XRBs. These include:
\begin{enumerate}
\item As the radio spectrum originates from synchrotron photons,
  energetic electrons exist in the jet. These electrons must upscatter
  low energy photons.
\item From a theoretical perspective, models in which the dominant
  contribution is IC emission from the inflow do not well connect to
  the need for strong magnetic fields required in leading
  jet-launching models discussed above (though a few recent accretion models
  may overcame this problem; see \cite{Ferreira+06, FM09, Bu+09,
    Oda+10, Petrucci+10}).
\item Detection of X-ray emitting blobs propagating outward in the
  inner regions of jets in several microquasars [\cite{Corbel+02}]
  indicate that part of the radiation in these objects is from the jet (or
  interaction of the jet with the ambient medium), and not all of it originates from the
  accretion flow.
\item Finally, the very high energy emission ($\gsim 100 $~GeV)
  observed in several microquasars (or microquasar candidates)
  [\cite{Aharonian+05, Albert+06, Albert+07a}] is difficult to explain
  in disk models.
\end{enumerate}

\subsection{Origin of the seed photons.}

As particles acceleration were discussed in \S\ref{sec:rad1} above,
let us focus on the origin of the seed photons for IC scattering.

{\bf Synchrotron self Compton.}  A natural source of seed photons are the synchrotron
photons emitted by the energetic electrons, namely SSC.  As long as the
scattering is in the Thomson regime, namely the energy of the
upscattered photons is much less than the energy of the incoming
electron, the outgoing photon energy is $\varepsilon_{\rm out} \simeq
4 \gamma_{el}^2 \varepsilon_{\rm in}$, where $\gamma_{el}$ is the Lorentz factor
of the electron [e.g., \cite{RL79}].  In this case, the spectral shape
of SSC emission from a power law distribution of energetic electrons
is {\bf similar} to that of synchrotron emission discussed above [see,
  e.g., \cite{SE01}]. It is characterized by four break
frequencies. If $\nu_m < \nu_c$, the values of these break frequencies
are $\nu_{SSA}^{IC} \simeq 4 \gamma_m^2 \nu_{SSA}$, $\nu_{m}^{IC}
\simeq 4 \gamma_m^2 \nu_{m}$, and $\nu_{c}^{IC} \simeq 4 \gamma_c^2
\nu_{c}$. If $\nu_c < \nu_m$, spectral break corresponding to the self
absorption occurs at $\nu_{SSA}^{IC} \simeq 4 \gamma_c^2 \nu_{SSA}$,
while the energy of the other breaks is not changed.

Thus, for $\nu_m < \nu_c$ the peak of the IC spectrum is at
$\nu_m^{IC}$ while for $\nu_c < \nu_m$ it is at $\nu_c^{IC}$, both
naturally extend up and above the MeV range.  The ratio of the IC and
synchrotron energy fluxes is given by $(\nu F_\nu)_{\rm peak, IC} /
(\nu F_\nu)_{\rm peak, syn} = Y$, where $Y = (4/3) \gamma_{el}^2 \tau$
is the Compton $Y$ parameter and $\tau$ is the optical depth. 

\begin{figure*}
 \includegraphics[width=0.9\textwidth]{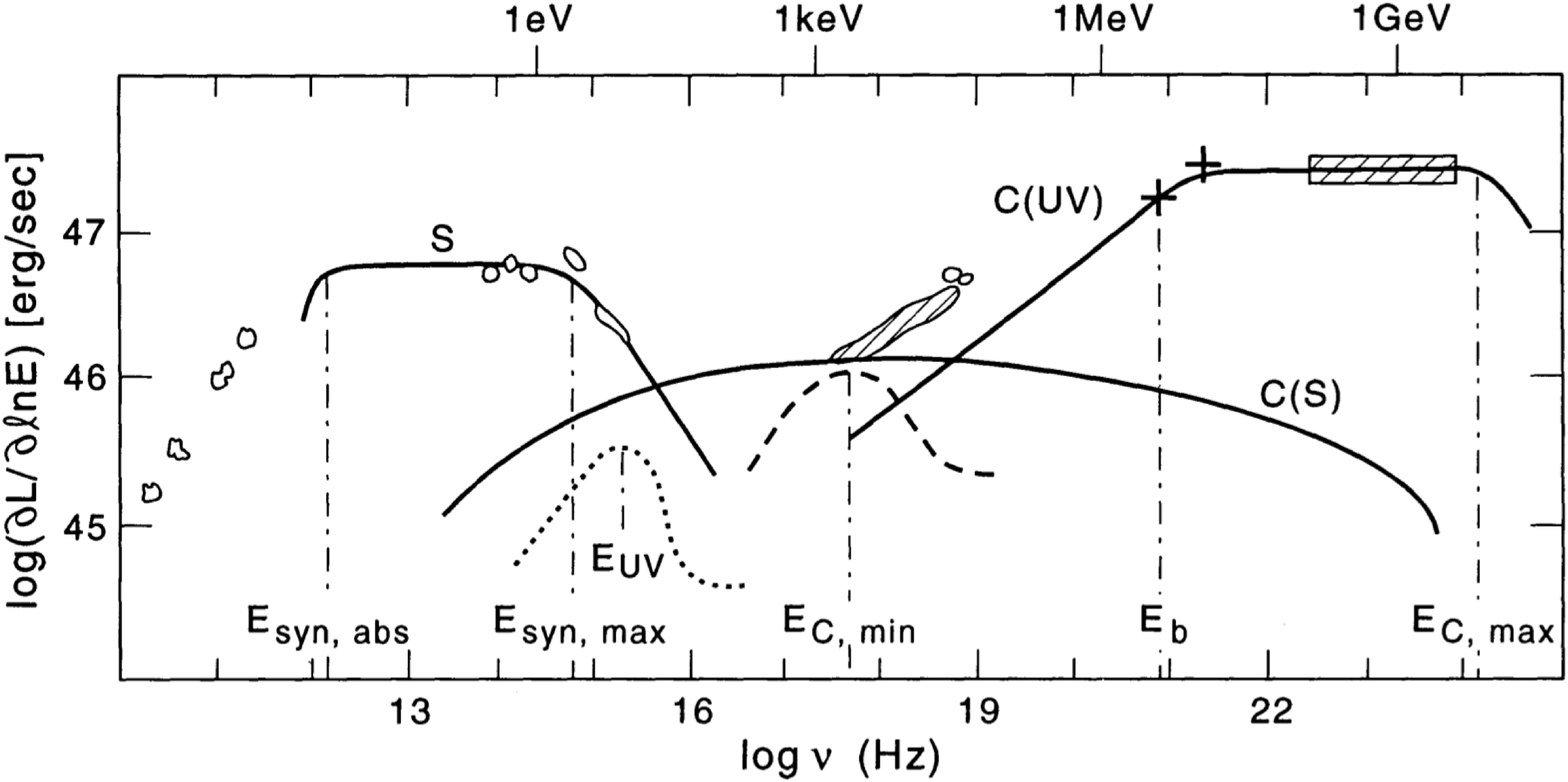}
\caption{Broad band spectrum of the AGN 3C 279, taken from
  \cite{SBR94}.  Solid lines mark the fitted model spectral
  components: (S) - synchrotron, C(S) - SSC, C(UV) - Comptonization of
  diffusive UV radiation. For further details as well as references
  about the data points, see \cite{SBR94}.  
}
\label{fig:broad1}      
\end{figure*}

In AGN jets two distinct, broad spectra components are observed. The
low energy component is peaking at the sub-mm to IR regime (in FSRQ)
or in the UV / X-rays (in high-frequency peaked BL Lacs, or HBLs). The
high energy component peaks at the MeV energies in FSRQs and GeV
energies in HBLs [e.g., \cite{Fossati+98, Donato+01, Sambruna+04,
    Levinson06} and references therein].  An example of this spectra
is presented in Figure \ref{fig:broad1}, taken from \cite{SBR94}.  It
is therefore natural to attribute the peak at the radio band to
synchrotron emission, while that at the X-ray band to IC, as is done
by many authors [\cite{Konigl81, MG85, MGC92, Hartman+01, FDB08,
    GTG09, Ghisellini+10} and many more]. In fact, lacking good
theoretical knowledge of the electron number density, hence of the
optical depth, very often it is being determined by fitting the ratio
of IC peak flux to the synchrotron peak flux. In recent years, similar
fitting was done to the X-ray spectra in XRBs [e.g., \cite{GBD06,
    Gallo+07, Migliari+07, Maitra+09}].

{\bf External seed photons.}  In addition to SSC, there are other
sources of seed photons. As the spectral shape of the resulting IC
emission depends on the spectral shape of the incoming photons, broad
band fitting of the spectra are required to determine which is the
dominant field. One natural source is the photon field created by the
accretion disk [\cite{BS87, Dermer+92, DS93, BL95}].  In XRBs, photons
from the companion star can also serve as seed photons for IC
scattering [\cite{DB06}].

Alternative source of photons is reprocessing of disk emission by the
surrounding material, such as the broad emission line region in AGNs
[\cite{SBR94, Dermer+97}; see Figure \ref{fig:broad1}]. Additional
suggestions for seed photons include reprocessing of the synchrotron
emitted photons from the jet itself by the surrounding medium before
being IC scattered [\cite{GM96}], infrared emission from circumnuclear
dust [\cite{Bla+00}] or synchrotron radiation from other regions along
the jet itself [\cite{GK03}]. Clearly, these models require additional
assumptions about the environment and/or the material that acts to
reprocess the original emission. The addition of degrees of freedom
with respect to the synchrotron-SSC model enables much better fits to
existing broad band data at the price of more complex modeling of the
environment.

In nearby objects such as Cen~A, the extended giant radio lobes can be
spatially resolved. Two distinguished spectral components, one at the
radio band [\cite{Hardcastle+09}] and one at the X-ray and/or
$\gamma$-rays [\cite{Feigelson+95, KS05, Croston+05, Abdo+10b}] are
detected in these lobes. While the radio spectrum is naturally
attributed to synchrotron emission, at such large distances from the
core (typically hundreds of kpc) it is insufficient to provide enough
photons to explain the X- and $\gamma$-ray flux observed. Instead,
this is attributed to IC emission of the cosmic microwave background
(CMB) or extra-galactic background (EBL) light [\cite{Tavecchio+00,
    CGC01}], whos' spectra are well known [\cite{Georg+08,
    FRD10}]. This scenario has a great advantage, as it enables
decoupling of the electrons distribution and the magnetic field.
First, the electron distribution is inferred from the IC spectra
and the known seed photon field, and at a second step the magnetic
field strength is inferred from the synchrotron spectrum. This
separation thus enables to infer the values of the magnetic fields in
these regimes, which is found to be close to equipartition. Since
these values are much larger than can be achieved by Poynting-flux
conservations from the core, as well as higher than the (compressed)
external field, these results point towards magnetic field generation
in shock waves, as discussed above.  Moreover, analyzing the spectra
enables to show that the conditions in these lobes enable
acceleration of particles to ultra-high energies [\cite{PL12}].

\subsection{Separation between disk and jet photons}

As much as inferring the origin of seed photons at large distances along
the jet is not easy, close to the jet base the situation is far more
complicated. Separating jet-based emission models from disk-based
emission models is a very difficult task, as both disk-based and
jet-based models can produce good fits to the data
[\cite{Markoff+05}].

On the one hand, there are some indirect evidence based on correlation
between emission at the X-ray band and lower energy bands (radio, IR
and optic) for jet-dominated X-ray emission [\cite{YYH09,
    Russell+10}]. This interpretation is strengthen by extrapolation
of the spectral energy distribution (SED) above the turn over at mid-
IR bands [\cite{Gandhi+11, Rahoui+11, Rahoui+12, Russell+13,
    Corbel+13}]. An independent support comes from polarization
measurements by {\it INTEGRAL} satellite [\cite{Laurent+11,
    Jourdain+12}], which show strong polarization above 400 keV in Cyg
X-1, hinting towards jet origin in these energies.

In spite of these indications, it should be stressed that as of today,
disk-based models for the X-ray emission are much more developed, and
are favored by a central part of the community. As these models are
thoroughly discussed in other chapters of this book, I will only
briefly mention some aspects here. In these models, many of the X-ray
properties are explained by Comptonization in hot accretion flows,
including detailed X-ray spectral shape [e.g., \cite{Sob+12, QL13}],
spectral evolution during state transition [\cite{DelSanto+13}], and
many timing properties of X-ray variability [\cite{Kotov+01, ID11,
    ID12}]. Moreover, detailed fitting of X-ray spectra with the
jet models require, in some cases, optical depth of $\tau \sim 2-3$
[\cite{MB09, PV09, DBMJ10}], which put strong constraints on the jet
kinetic power [\cite{MBF09}]. Further critical  discussions about jet
vs. disk models can be found in \cite{PZ03, Zdziarski+03, Maccarone05,
  Veledina+13} as well in other chapters of this book.

\begin{figure*}
 \includegraphics[width=0.9\textwidth]{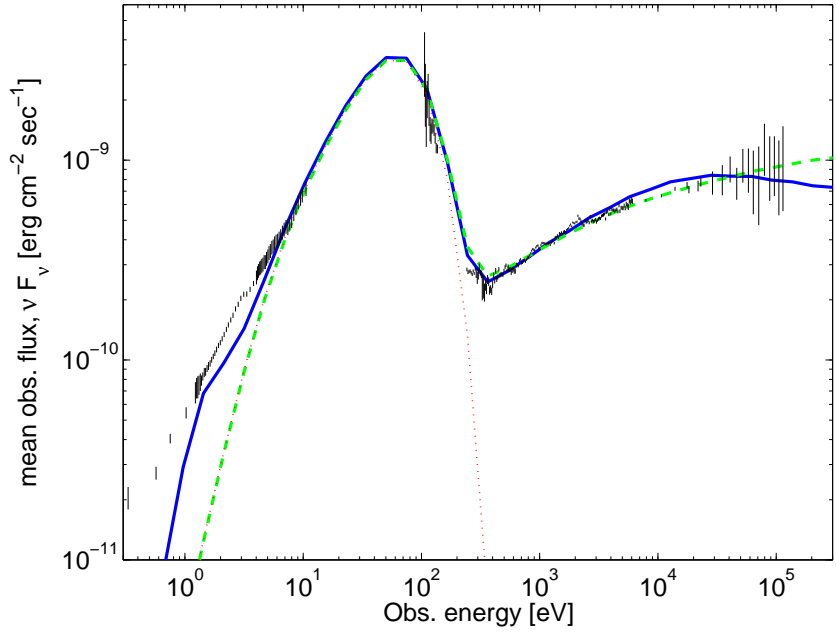}
\caption{Fitting the 2000 outburst of XTE J1118+480 (taken from
  \cite{PM12}). The
  solid (blue) curve represent a model in which synchrotron emission
  is the main source of radiation, while the dashed (green) represents
  an IC-dominated model. Both models provide good fits to the data,
  and are consistent with emission from electrons in the inner part of
  the jet. For further details see \cite{PM12}.  }
\label{fig:PM12}      
\end{figure*}

One difference between the two scenarios is that while electrons in
the inflow are expected to be continuously heated as they spiral in,
it is possible that once they enter the jet region they are no longer
heated. As they propagate outwards inside the jet, both the radiation
field and the magnetic field decay, and thus cooling of the electrons
is suppressed. During their initial propagation outward, they do
though radiatively cool very rapidly.  For rapidly cooling electrons,
both synchrotron emission discussed above and Compton scattering
produce the same spectrum: $F_\nu \propto \nu^{-1/2}$ in the range
$\nu_c < \nu < \nu_m$. This spectrum is consistent with the X-ray
spectra observed in many outbursts in XRBs [e.g., \cite{Hynes+00,
    Esin+01, Homan+05, Joinet+05}] \footnote{Jet dominated models are
  expected in XRBs during the low/ hard state, where $L \sim 1\%
  L_{Edd}$. At higher luminosities, disk contribution is expected, and
  the spectral slope varies; the luminosity-dependence of the spectral
  index can be found in \cite{WG08}.}. Thus, by identifying the break
frequencies seen in the spectra with $\nu_m$ and $\nu_c$, it is
possible to constraint the physical parameters from the emitting
region. In particular, this analysis may enable to discriminate
between emission from the inner parts of the jet (in which the
electrons reside only a short time), and emission from the inflow,
which is expected to last over a longer period, during the
spiral-in. Such an analysis was carried by \cite{PM12}, and one of its
results is presented in Figure \ref{fig:PM12}.\footnote{Data taken from \cite{McClintock+01}; Note,
    though, that a different analysis of BeppoSAX [\cite{Frontera+01}]
    and Chandra [\cite{Reis+09}] data done in the context of disk
    models, resulted in a somewhat softer slope below 2 keV.}

\section{Hadronic contribution to the high energy spectra}

The uncertainty in the nature of the acceleration process implies that
protons may be accelerated as well inside the jets. Once accelerated
to high energies, protons contribute to the observed
spectrum. Although synchrotron emission and Compton scattering are
suppressed with respect to emission from leptons due to the much
smaller cross section, protons may still have a significant
contribution to the high energy emission. First, if the acceleration
process acts in such a way that most of the energy is deposited in
accelerated protons, it is possible that synchrotron emission from
these protons have a significant contribution to the high energy (X-
and $\gamma$-ray) flux [\cite{Aharonian00}]. Second, energetic protons
can deposit their energy by photo-meson production, $p \gamma
\rightarrow n + \pi^+, p + \pi^0$ [\cite{MB92, Mannheim93}]. The
created $\pi$ mesons are unstable; the $\pi^+$ can radiate synchrotron
emission before decaying into $\mu^+ + \nu_\mu \rightarrow e^+ + \nu_e
+ {\bar \nu_e} + \nu_\mu$, while the $\pi^0$ decays into a pair of
energetic photons. These particles thus produce a high energy
electromagnetic cascade, as the created photons are energetic enough
to produce a pair of electron-positron, $\gamma + \gamma \rightarrow
e^\pm$. Emission from these secondaries may thus be responsible for
the high energy (up to TeV) emission seen in AGNs (blazars)
[\cite{RM98, Mucke+03, Murase+12}], as well as in GRBs [\cite{PW04, PW05b}]
and XRBs [\cite{Romero+05}].

\begin{figure*}
 \includegraphics[width=0.9\textwidth]{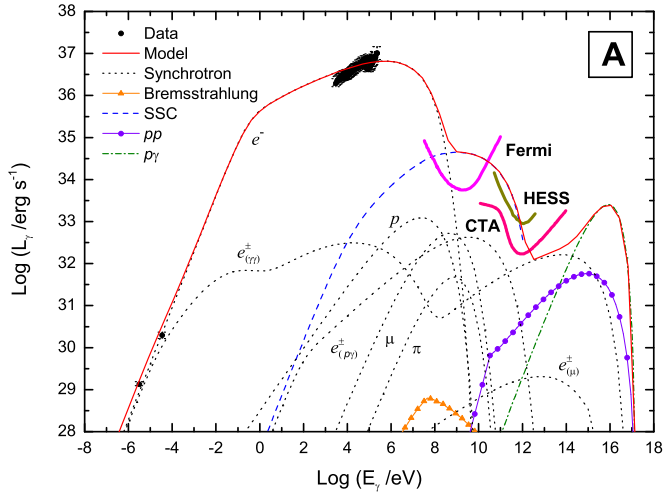}
\caption{Fitting the broad band spectra of the 1997 outburst of GX
  339-4 by a leptonic/hadronic model of \cite{VR10}. Contribution from
  the various processes are marked: generally, protons contribution is
  expected mainly at high energies, X- and $\gamma$-rays. These fits
  are used to infer the uncertain values of the magnetic field as well
  as the electron/proton distribution. For details see \cite{VR10}. }
\label{fig:VR10}      
\end{figure*}

In addition to photomeson production, protons can interact with
photons by photopair production ($p + \gamma \rightarrow p + e^\pm$),
and with other protons through proton-proton (pp) collisions,
producing pions and Kaons [\cite{KPW06}]. The rate of these
interactions depend of course on the ambient photon field, as well as
the uncertain distribution of the energetic protons. Combined
leptonic/hadronic models that explain the radio emission in AGNs and
XRBs as due to synchrotron radiation from electrons and the high
energy (up to TeV) emission as due to hadronic-originated cascade
exist [e.g., \cite{VR10}; see Figure \ref{fig:VR10}]. Although, as can
be seen from Figure \ref{fig:VR10}, these models are still lagging
behind disk models, and can thus currently can still only provide
approximate fits to the observed spectra. These models suffer two main
drawback: First, the inherent uncertainty in the knowledge of the
accelerated proton distribution. Second, calculating the evolution of
the high energy electromagnetic cascade is extremely difficult, due to
the non-linearity of the process, and the fact that it is very rapid,
namely, many orders of magnitude shorter than the dynamical
time. Thus, it is numerically challenging. While in recent years
models of cascade evolution in GRB environment exist [\cite{PW05}],
this field is still at its infancy.

An interesting consequence of hadronic models, is that if indeed
protons are accelerated to high energies, among the secondaries
produced are high energy neutrinos [for a review about neutrino
  production in AGN jets see \cite{GHS95}; for neutrino production in
  XRB jets, see \cite{LW01, Chris+06, Zhang_JF+10}] . Thus, such neutrinos
- if detected - would be a direct proof of proton acceleration in these
environments.

\section{Temporal information}

\subsection{XRBs: temporal correlation between different spectral bands}

Although the basic radiative processes are well known, the emitted
spectra from jets are very complex, due to the complex nature of these
systems. Emission originates from both the accretion flow, different
regions along the jet where the physical conditions vary, as well as
external photons that can be reprocessed (IC scattered) from particles
along the jet. The physical conditions along the jet, such as the
magnetic field and particle distribution and their connection to the
physical conditions in the inner parts of the inflow are uncertain.
It is thus not surprising that the observed spectra can be interpreted
in more than one way, and that plethora of models aimed at explaining
the broad band spectra exist. An in depth discussion in some of the
models appear in the chapters authored by Fender, Gallo, Casella and
K\"ording in this book.

Thus, in order to obtain a full picture additional information is
needed. In XRBs, a natural source of information is temporal analysis,
since the emission pattern conveniently changes over time scale of
$\lsim$~months. As the emission changes with time, correlation between
emission from the inflow and the jet at different times (the different
``states'') is established. Such correlation is the switch off of jet
radio emission in the high/soft state [\cite{Tananbaum+72,
    Fender+99b}]. Others are the correlation found between the X-ray
luminosity and the radio luminosity [\cite{HHCS98, Corbel+00,
    Corbel+03, Gallo+03}], which are found to scale as $L_R \propto
L_X^{0.7}$. However, recently, it was shown that the system H1743-322
follow a different correlation, $L_R \propto L_X^{1.4}$
[\cite{Coriat+11, Gallo+12}]. Other correlations are found at
different wavebands: between the X and near infrared (NIR)
[\cite{Russell+06, Coriat+09, Casella+10}], and radio - optic and
X-rays [\cite{Kanbach+01, Gandhi+10, Cadolle-Bel+11}].

The wealth of emitting zones and radiative processes enables to
interpret the observed correlations in various ways. One type of
models explore the obvious (yet uncertain) connection between the
properties of the inner parts of the accretion flow and the jet [e.g.,
  \cite{Markoff+03, HS03}]. Other ideas include the obvious connection
between the synchrotron radiation and IC scattering by the same
population of electrons [\cite{Giannios05, Veledina+11}; see details
  in the chapter by Poutannen in this book], as well as correlation
between (synchrotron) emission by the same electrons as they propagate
along the jet thereby occupying different regions in the jet at
different times [\cite{Casella+10}].

Existence of different emission zones reflects the complex internal
dynamics of the outflow. For non-steady outflow, shock waves naturally
develop when two ``blobs'', or shells of plasma collide. This happens
once the ejection of a slower plasma blob is followed by ejection of
faster moving one. Once the blobs collide, two shock waves are formed,
propagating into both plasmas. By heating (and possibly accelerating
to high energies) the particles, these shock waves are the initial
source of radiation. This scenario was invoked to explain the complex
lightcurve seen during GRB prompt emission [\cite{RM94, DM98}]. In
recent years, similar ideas were studied in the context of emission
from XRBs [\cite{KSS00, Jamil+10, Malzac13}], and TDEs [\cite{GM11};
  although a structured jet model was suggested by \cite{LPL13}].

This idea, though, is incomplete: currently, the internal shock model
is lacking a predictive power about the radii at which the collisions,
hence the energy dissipation takes place - these are determined by the
initial conditions.  Thus, overall, my personal opinion is that
understanding the nature of the correlations observed is at its
infancy, and that this field is a very promising path to take. Future
models will inevitably combine both dynamical models and radiative
models, which will mature in the coming years.

\subsection{Flaring activities in AGNs}

In AGNs (blazars), flaring activity is observed in the X- and
$\gamma-$rays up to the highest energies, at the TeV band. This is
often observed on a very short time scales, of the order of hours and
in some cases even minutes [\cite{Kniffen+93, Buckley+96,
    aharonian+07, albert+07, Aleksic+11}].  Radio observations showed
that radio outburst seem to follow the $\gamma$-ray flares
[\cite{Reich+93, Zhang+94}]. While significant variability in the
optical band is observed as well, its correlation with the variability
in the $\gamma$-band is not fully clear [\cite{Wehrle+98, Palma+11}].

The main implication of this rapid variability in the flux is
constraining the size of the emitting region and the bulk motion
Lorentz factor. An observed variability time $\Delta t^{\rm ob}$
implies that the size of the emitting region cannot exceed
\beq
r \leq r_{\rm var} \approx {\Gamma c \Delta t^{\rm ob} \mathcal{D} \over 1 + z},
\label{eq:var}
\eeq
where $z$ is the redshift, and $\Gamma$ is the Lorentz factor
associated with the bulk motion [\cite{GM96}].  On the other hand, the
fact that TeV photons are observed implies that the optical depth to
pair production with the low energy photons in the plasma cannot
exceed unity. Thus, the emitting region cannot be too
compact. Combined together, these two constraints imply high bulk
Lorentz factor (e.g., in PKS 2155-304, $\Gamma \gsim 50$ was inferred
by \cite{BFR08}).  The exact value of the constraint on the emitting
region thus depend on the variability time, as well as the photon
field. The variability itself reflects changing conditions within the
outflow, e.g., due to the existence of internal shock waves
[\cite{Spada+01}]. The fact that the constraints found on $\Gamma$ in
PKS 2155-40 were found to be inconsistent with direct measurements,
have led \cite{GUB09} to suggest a jet within a jet model for the high
energy emission. In a more general form, this can be viewed as an
indication for an internal structure within the jets.

\section{Jet power}

Estimating the total deposited energy (or power) in astronomical jets
is a very tricky task. The complexity of the problem is most easily
understood if one considers the different episodes of energy transfer
in these systems.  First, there is the {\it kinetic} energy associated
with the bulk motion of particles inside the jet. Naturally, this is
some fraction of the gravitational energy of the inflowing material in the
accretion disk. Theoretical determination of this fraction is possible
only after the theory of jet production is fully understood. Until
then, it can only be estimated from observations.

The observed radiation, in turn, reveals only a small fraction of this
energy. Following jet launching, the second energy transfer occurs at
a certain location(s) along the jet, where particles are accelerated
to high energies. This acceleration must occur on the expense of (part
of) the bulk motion kinetic energy, but possibly also due to magnetic
reconnection - in which case it is at the expense of magnetic
energy. Finally, the accelerated particles radiate {\it some fraction}
of their energy as photons, producing the observed signal.  Thus,
direct observation of the photon signal reveals only an unknown
fraction - likely a small fraction, of the kinetic energy initially
given to the particles inside the jets.

Estimating the kinetic jet power is thus difficult, and rely on
several assumptions. For example, \cite{RS91} estimated the average
kinetic power of jets in blazars by dividing the total energy stored
in the form of electrons and the magnetic field energy in the radio
lobes (as deduced from synchrotron theory and the equipartition
assumption) by the lobe age, which was computed from spectral aging or
expansion velocity arguments. Similarly, \cite{CF93} estimated the jet
power of blazer jets using the framework of the standard synchrotron
self-Compton theory.  As explained above, these works suffer from
substantial uncertainties, due to the various underlying assumptions
needed.

In an alternative approach, \cite{Allen+06, BBC08} estimated the jet
kinetic power by estimating the mechanical work, $PdV$ required to
inflate the observed giant X-ray cavities. Since here too there are
uncertainties in estimating the size of these cavities, these
translate into uncertainties in the jet power.  This will further be
discussed in the chapter of this book authored by Heinz.

These works found a strong correlation between the estimated jet power
and the disk luminosity. Such a correlation is expected in the leading
mechanisms for jet production. As material from the jet originates
from the disk, such correlations are of no surprise. Additional clue
may come from a correlation between the jet power (as estimated from
the radio flux) and spin of the black hole, as recently reported
[\cite{NM12, Steiner+13}]. While this result is still debatable (see
\cite{RGF13}), if confirmed it may serve as a strong clue for the
mechanism that launches jets in nature. See further discussions in the
chapters by McClintock, Narayan, Fender and Gallo in this book.

Thus, while various models that estimate the kinetic power of jets
exist, they all suffer from uncertainties, caused both by
uncertainties in the measurements, and also by the need to rely on
uncertain emission models. I thus view this subject as one which is far
from being matured, and will be further developed in the near future.

\section{Summary and conclusions}

In this chapter, I reviewed some of the basic radiative mechanisms that
produce the broad band emission seen in astronomical jets. Due to the broad nature
of this subject, I focused on XRBs and AGNs (mainly blazars). The main
radiative processes considered are synchrotron emission, SSC, Compton
scattering of external, or reprocessed photons, and hadronic
contribution, via proton-synchrotron emission and electromagnetic
cascade caused by secondaries produced by proton-photon (and to a
lesser extent, proton-proton) interactions.

Although each of these processes is well understood, the changing
conditions inside the jets lead to complex observed spectra. This
leads to the fact that inspite a wealth of broad-band data, no single
model is commonly accepted. On the contrary, as discussed here, the
same data can be interpreted in more than one way. Thus, the main
``take away'' massage from this chapter, is that modeling emission
from jets is one of the most challenging tasks.

The questions that need to be addressed when studying emission from
jets extend far beyond the realm of the radiative processes involved,
and require addressing questions in basic physics and astronomy.
Broadly speaking, in order to fully understand the emission, one needs
to understand:
\begin{enumerate}
\item The connection between disk and jet, and the mechanism that
  leads to jet launching. 
\item The varying physical conditions in different regions inside the
  jet, such as the magnetic field along the jet.
\item The jet composition that governs the contribution of leptons and
  hadrons to the observed spectra.
\item The nature and details of the acceleration mechanisms that
  determine the energy distribution of energetic particles in
  different parts of the jet.
\item The internal (synchrotron) and external (accretion disk,
  companion star, CMB, etc.) photon fields that serve as seed photons
  to scattering by jet material.
\item The geometry of the jets, including velocity profile and its
  angle towards the observer, that determine the different scattered
  field, as well as the Doppler boost.
\item The dynamics of material inside the jet, that determines the
  spatial distribution of the radiating particles and their temporal
  evolution.
\end{enumerate}

Addressing each of these questions is a task so challenging by itself,
that despite decades of research and numerous works (unfortunately,
only very few could be mentioned here) we still have only clues, but
no definite answer to any of them. Moreover, these questions, while
can be addressed separately, should be addressed in the context of the
different environments in which jets are observed - XRBs, AGNs, GRBs
and recently also TDEs. Thus, full answer to all these questions is
not expected any time in the near future. However, the wealth of
current and future data - both spectral, temporal and spatial data,
ensures that there is plenty of room for new ideas in the coming
years.

\begin{acknowledgements}
I would like to thank Paul Callanan, Piergiorgio Casella, Amir Levinson and Abraham Loeb for providing useful comments on this manuscript.
\end{acknowledgements}



\end{document}